\documentstyle[aps,preprint]{revtex}
\newcommand{\bb}{\begin{eqnarray}}
\newcommand{\ee}{\end{eqnarray}}
\begin{document}
\title{{Temperatures of  extremal black holes}}
\author{Amit Ghosh\thanks{e-mail amit@saha.ernet.in}
and P. Mitra\thanks{e-mail mitra@saha.ernet.in}}
\address{Saha Institute of Nuclear Physics\\
Block AF, Bidhannagar\\
Calcutta 700 064, INDIA}
\date{gr-qc/9507032}
\maketitle
\begin{abstract}
The temperature of an extremal Reissner - Nordstrom black hole is not
restricted by the requirement of absence of a conical singularity.
It is demonstrated how Kruskal-like coordinates may be constructed
corresponding to any temperature whatsoever.
A recently discovered stringy extremal black hole which apparently has an
infinite temperature is also shown to have its temperature unrestricted by
conical singularity arguments.
\end{abstract}

\bigskip

A classical black hole has a  horizon  beyond which  nothing
can leak out. This suggests that it can be assigned a zero
temperature. But the relation between the area
of the horizon and the mass and other parameters like the charge
indicates a close similarity \cite{BCH} with
the  thermodynamical laws, thus allowing the definition of a temperature.
This analogy was understood as being of quantum origin and made  quantitative
after  the  discovery of Hawking radiation \cite{Hawk}. The associated
Hawking temperature vanishes only in the classical limit.
The thermodynamics of black holes has been extensively studied since then.

Most of the studies were first made for the simplest kind of black
hole, {\it viz.}, the Schwarzschild spacetime.
Of more recent interest is the case of the so-called extremal black holes which
have peculiarities not always present in the corresponding
non-extremal cases \cite{Pres,GM}.
For extremal Reissner - Nordstrom black holes,
the na\"{\i}vely defined temperature is zero, but
the {\it area}, which is usually thought of as the entropy, is nonzero.
For extremal  dilatonic  black  holes,  where  the temperature is
{\it not} zero, the area vanishes.

In this note we shall reexamine the temperature of an
extremal Reissner - Nordstrom black hole, for which
the definition in terms of the surface gravity  leads to a zero temperature.
We discuss the conical singularity approach in detail: there is {\it no}
conical
singularity in the extremal case, so that there is {\it no} constraint on the
temperature from this point of view. Thereafter we consider Kruskal-like
coordinates and show that they may be constructed for an arbitrary temperature.

We also comment on some stringy black holes. One of them has an infinite
temperature in the surface gravity approach. We show however that there is no
conical singularity in this case.

\section*{Reissner - Nordstrom black holes}

The metric of the Reissner - Nordstrom spacetime is given by
\begin{equation}
ds^2=-(1-{2M\over r}+{Q^2\over r^2})dt^2+ (1-{2M\over r}+{Q^2\over r^2})
^{-1}dr^2 +r^2(d\theta^2+\sin^2\theta d\phi^2)
\end{equation}
in general, with $M$ and $Q$ denoting the mass and the charge
respectively. There are apparent singularities at
\begin{equation}
r_\pm=M\pm\sqrt{M^2-Q^2}
\end{equation}
provided $M\ge Q$. Cosmic censorship dictates that this
inequality holds and then there is a horizon at $r_+$.
The limiting case when $Q=M$ and therefore $r_+=r_-(=M)$ is referred to as
the extremal case. Whereas the case with $Q<M$ is qualitatively similar to a
Schwarzschild black hole, it is clear that the horizon in the extremal case
will behave differently: the metric singularity becomes stronger here.

The best known method of calculating the temperature of a black hole is
through the relation with surface gravity. To
distinguish this  temperature from those arising in other
approaches, we may call it the Unruh temperature. Here
\bb
T&=&{1\over 2\pi}{1\over\sqrt{g_{rr}}}\left.{d\sqrt{-g_{tt}}\over dr}
\right|_{r=r_+}
\nonumber\\
&=&{1\over 2\pi}\sqrt{(1-{r_+\over r})(1-{r_-\over r})}\left.
{d\sqrt{(1-{r_+\over r})(1-{r_-\over r})}\over dr}\right|_{r=r_+}\nonumber\\
&=&{r_+-r_-\over 4\pi r_+^2}.
\ee
Now this expression vanishes in the extremal limit where $r_+=r_-$. As
the properties of extremal black holes are qualitatively different from
those of non-extremal ones, it is better to calculate the temperature afresh
in the extremal case. One finds
\bb
T&=&{1\over 2\pi}{1\over\sqrt{g_{rr}}}\left.{d\sqrt{-g_{tt}}\over dr}
\right|_{r=M} \nonumber\\&=&{1\over 2\pi}(1-{M\over r})
\left.{d(1-{M\over r})\over dr}\right|_{r=M} =0.
\ee
Thus the result is the same as is obtained through the limiting procedure. This
means that the surface gravity is continuous in this limit.

Next we consider the question of a conical singularity on passing to
imaginary time. The metric
\bb
ds^2= dr^2 + r^2d\theta^2,\label{pol}
\ee
which describes the flat Euclidean metric in polar variables, can be
supposed to describe distances on the surface of a cone. The cone has a
singularity at its tip $r=0$, except in the limiting case when the cone
opens out as a plane. In this situation $\theta$ has a periodicity $2\pi$,
so one may say that the conical singularity is avoided by making $\theta$ an
angular variable with this period. This is relevant for black holes because
such a singularity tends to arise in the Schwarzschild and in the
non-extremal cases. In the latter case, one passes to imaginary time and
writes the metric for a constant $\theta,\phi$ surface as
\bb
ds^2&=&(1-{r_+\over r})(1-{r_-\over r})dt^2 + {dr^2\over
(1-{r_+\over r})(1-{r_-\over r})}\nonumber\\
&=&\Omega(\rho)(d\rho^2 + \rho^2d\tau^2),
\ee
where $\tau=\alpha t$ with the constant $\alpha$ so chosen as to make
the conformal factor $\Omega$
finite at the horizon. For consistency, one requires
\bb
\rho= e^{\alpha r_*},
\ee
where $r_*$ is defined by
\bb
dr_*={dr\over (1-{r_+\over r})(1-{r_-\over r})}.
\ee
Near the horizon {\it in the non-extremal case},
\bb
r_*\approx{r_+^2\over r_+-r_-}\log (r-r_+),\label{log}
\ee
so that
\bb
\rho\approx (r-r_+)^{\alpha r_+^2/(r_+-r_-)},
\ee
which implies that $\rho$ vanishes at the horizon, and
\bb
\Omega={(1-{r_+\over r})(1-{r_-\over r})\over\alpha^2\rho^2}
\ee
can be made finite at the horizon by making $\rho^2$ vanish linearly as $r\to
r_+$, {\it i.e.,} by choosing $\alpha$ to satisfy
\bb
{\alpha r_+^2\over r_+-r_-}={1\over 2}.\label{alpha}
\ee
Now for the conical singularity to be avoided, one must have
a periodicity of $2\pi$ for $\tau$, {\it i.e.}, a periodicity for
$t$ given by ${2\pi\over\alpha}$. This corresponds to a temperature
\bb
T={\alpha\over 2\pi}= {r_+-r_-\over 4\pi r_+^2},\label{T}
\ee
which is the standard non-extremal result given above. Thus for a non-extremal
black hole, what may be called the {\it conical} temperature
agrees with the Unruh temperature.

In the extremal case, things are very different. Near the horizon, now,
\bb
r_*\approx -{M^2\over r-M},\label{lin}
\ee
so that $\rho$ has an essential singularity as $r\to r_+$, and there is
no value of the constant $\alpha$ which can make
the conformal factor $\Omega$ regular at the horizon. This
simply means that the extremal metric is not of the form (\ref{pol})
and there is no question of a {\it conical} singularity. Consequently,
the derivation of a {\it conical} temperature fails. As far as this approach is
concerned, one may say that the temperature has no specific constraint of
periodicity to satisfy, and is therefore arbitrary in this extremal case
\cite{HHR,moretti}.

We come finally to the question of an analogue of Kruskal coordinates for these
black holes: the new coordinates have to be such that the metric components are
nonsingular at the horizon. This topic may seem unconnected to our main
interest, which is the temperature, but the Kruskal vacuum plays an important
r\^{o}le in the theory of Hawking radiation and hence in the idea of the
Hawking
temperature \cite{Hawk}. One first writes the metric of a surface with
constant $\theta,\phi$ in the double null form
\bb
ds^2&=&-(1-{r_+\over r})(1-{r_-\over r})(dt^2 - dr_*^2)
\nonumber\\&=&-(1-{r_+\over r})(1-{r_-\over r})dvdw
\ee
with
\bb
v=t-r_*,~~~w=t+r_*.
\ee
Here $r$ is understood to be implicitly defined by using $w-v=2r_*$ and
the relation between $r_*$ and $r$. One passes to new null coordinates $\bar
v$ and $\bar w$ defined by
\bb
v=f(\bar v),~~~w=g(\bar w),
\ee
with appropriate functions $f,g$. The metric becomes
\bb
ds^2
=-(1-{r_+\over r})(1-{r_-\over r}){df\over d\bar v}{dg\over d\bar w}
d\bar{v}d\bar{w}
\label{Q}\ee
and $r$ is understood to be determined implicitly by
\bb
g(\bar w)-f(\bar v)=2r_*.
\ee
The functions $f,g$ are to be chosen in such a way that the coefficient of
$d\bar{v}d\bar{w}$ in the right hand side of (\ref{Q}) is regular at the
horizon.

In the non-extremal case, the choice of the new coordinates is essentially the
same as in the Schwarzschild case. The horizon corresponds to $r_*\to -\infty$
and therefore either $v$ or $-w$ has to be infinite. The new coordinates are
defined by \cite{brill}
\bb
v=f(\bar v)=-{1\over\alpha}\log\bar v,~~~w=g(\bar w)
={1\over\alpha}\log\bar w,\label{krus}
\ee
with the constant $\alpha$ to be determined. This definition has
the result that one of the
new coordinates has to vanish at the horizon. If we consider
a point where $\bar w$ vanishes, we see that the factor ${dg\over d\bar w}
\propto {1\over\bar w}$ in (\ref{Q}) becomes infinite and may make the product
with $(1-{r_+\over r})$ finite. For this to happen, $\bar w$ must vanish
linearly as $r\to r_+$. Now, in the non-extremal case, (\ref{log}) indicates
that
\bb
{1\over\alpha}\log\bar w\approx{2r_+^2\over r_+-r_-}\log (r-r_+),
\ee
so that the condition for linearity of $\bar w$ is the same as (\ref{alpha}).
This fixation of $\alpha$ completes the definition of the new coordinates. We
have arranged the regularity of the metric components at vanishing $\bar w$,
but with this choice one can also check that there is no problem in the region
of vanishing $\bar v$.

Now we have to find the temperature. A time coordinate can be defined
in terms of the new null variables and a vacuum can be defined in terms of
this time. Green functions corresponding to this vacuum involve the new
coordinates $\bar v, \bar w$ (and $\theta,\phi$). If these Green functions can
be shown to have a periodicity in the original time $t$ after rotation to the
imaginary direction, the vacuum can be asserted to have a thermal character
\cite{davies}.
As $\bar v=e^{-\alpha v}, \bar w=e^{\alpha w}$ involve $\alpha t$, it is clear
that there is an imaginary period of length ${2\pi\over\alpha}$, which
corresponds to the temperature (\ref{T}) found above. In other words, for a
non-extremal Reissner - Nordstrom black hole, the standard Kruskal
temperature is the same as the Unruh and
conical temperatures. This is not surprising because the form of the metric
near the horizon of such a black hole is very similar to the Schwarzschild
form.

As in the conical approach, the case of the extremal black hole
is very different from the non-extremal cases. The
behaviour (\ref{lin}) of $r_*$ near the horizon is linear here
and the coefficient in (\ref{Q}) has an extra power of $(r-r_+)$.
So the logarithmic transformation does not
lead to metric components regular at the horizon. A transformation
that does has been known for a long time \cite{carter}:
\bb
v=f(\bar v)=M\tan\bar v,~~~w=g(\bar w) =M\cot\bar w.\label{tan}
\ee
It is not difficult to check the regularity of the
coefficient in (\ref{Q}). But the new coordinates do not exhibit any
periodicity in imaginary time, so that the vacuum corresponding to the time
defined by the above coordinates is {\it not thermal}.

One way of having a thermal vacuum would be to use coordinates as
in (\ref{krus})
but with $\alpha$ undetermined. This parameter cannot indeed be determined, as
no choice can make the metric components regular across the horizon. There is
an incompatibility between the twin requirements of regularity and
thermality. But a compromise can be made. Note that the question of regularity
concerns the region near the horizon. We have found that Green functions
cannot be arranged to be thermal in this region. But it should be
enough to have thermal behaviour at infinity. Instead of requiring that
the Green functions defined in the new vacua should have an imaginary
periodicity in $t$ {\it everywhere}, we shall impose this requirement
only for the region away from the horizon.

Thus, we require our new null coordinates $\bar v, \bar w$ to be of the form
(\ref{tan}) near the horizon ($v\to\infty$ or $w\to -\infty$) and of the form
(\ref{krus}) far away.
These two forms can be smoothly joined in the intermediate region in many
different ways, for instance by the equations
\bb
\bar v=\tan^{-1}[{v(1+e^{u-\sigma v})\over M}],~~~\bar w=
\cot^{-1}[{w(1+e^{u+\sigma w})\over M}],
\ee
where $\sigma$ is a positive constant akin to $\alpha$ in (\ref{krus})
and $u$ is a large positive constant. It is clear
that for large positive $v$ or large negative $w$, {\it i.e.,} near the
horizon, the new coordinates are very close to those given in (\ref{tan}),
but for negative $v$ and positive $w$,
the new coordinates have the exponential dependence on $v,w$
which is characteristic of thermal behaviour. The periodicity in imaginary
time is ${2\pi\over\sigma}$, so that the temperature is ${\sigma\over 2\pi}$.
This is of course arbitrary as the parameter $\sigma$ is free.

In conclusion, it may be repeated that the extremal
Reissner - Nordstrom black hole does not
have a unique temperature. The vacuum constructed using the time coordinate
corresponding to the smooth coordinates introduced in \cite{carter} is not
thermal. What we have shown is that alternative coordinates can be chosen for
which the vacuum is indeed thermal and the temperature involved is completely
arbitrary.

\section*{Stringy black holes}
Here we consider  extremal limits of some electrically charged
non-rotating black hole solutions of heterotic string theory compactified on a
6-dimensional torus \cite{sen}.
These black holes are characterized by the mass and a
28-dimensional charge vector (22  coming from the left hand sector
and the other 6 from the right hand sector of the theory). Among the various
extremal limits of these solutions we shall be interested only in those which
are space-time supersymmetric and hence saturate the Bogomol'nyi mass bound.
There are two such black holes which are described generically by the metric
\bb
ds^2=-R^{-1/2}r^2dt^2+R^{1/2}r^{-2}dr^2+R^{1/2}
(d\theta^2+\sin^2\theta d\phi^2)
\ee
and the dilaton
\bb
e^{\Phi}=r^2/R^{1/2}
\ee
where $R$ is a function of $r$ only. The solutions for the 28 gauge
fields are not of relevance in the following discussion.
The two cases are as follows.

{\bf a.} The first is characterized by the mass and charges
\bb
M={m_0\over 2}\cosh\alpha,\qquad \vec Q_L={m_0\over\sqrt 2}\sinh\alpha~
\vec n,\qquad \vec Q_R={m_0\over\sqrt 2}\cosh\alpha~\vec p,
\ee
which saturate the mass bound
\bb
M^2={1\over 2}\vec Q_R^2.
\ee
Here $m_0$ and $\alpha$ are two real parameters while $\vec n$ and $\vec p$ are
two unit vectors with 22 and 6 components respectively.
The function $R$ is given by $R=r^2(r^2+2m_0r\cosh\alpha+m_0^2)$.
In this case we can find a coordinate transformation which maps the
line element, after passage to imaginary time and
modulo the angular elements, to the form (\ref{pol}).
As in the non-extremal Reissner - Nordstrom case
there is a conical singularity at the tip
of the cone which lies on the horizon $r=0$. Regularity requires a particular
periodicity in the time coordinate which fixes the conical temperature
to be the same as the Unruh temperature of the black hole.

{\bf b.} The second case is more interesting because the  Unruh temperature
of this black hole blows up. Here
\bb
M={m_0\over 2},\qquad \vec Q_L={m_0\over\sqrt 2}~\vec n,\qquad
\vec Q_R={m_0\over\sqrt 2}~\vec p
\ee
with
\bb
M^2={1\over 2}\vec Q_R^2={1\over 2}\vec Q_L^2
\ee
and $R=r^2(r^2+2m_0r)$. One can rewrite the line element,
after passage to imaginary time and with the angular
part removed, in the form (\ref{pol}):
\bb
ds^2&=&R^{-1/2}r^2dt^2 + R^{1/2}r^{-2}dr^2 \nonumber\\
&=&\Omega(\rho)(d\rho^2 + \rho^2d\tau^2),
\ee
with $\tau=\alpha t$ as before and $\rho=\exp(\alpha r_*)$. In this case,
$r_*$ vanishes on the horizon (with an appropriate choice of the integration
constant involved), so that $\rho$ becomes unity, {\it i.e.,} the horizon does
{\it not} correspond to the tip of the cone. Furthermore, $\Omega$ is given by
\bb
\Omega^2={r^2\over \alpha^2r\sqrt{r^2+2m_0r}\rho^2},
\ee
which vanishes on the horizon. Thus the conical singularity does not arise
here.
Consequently, there is no constraint that can fix the parameter $\alpha$ and
$t$ can have any periodicity. As in the case of the extremal Reissner -
Nordstrom black hole, there is no definite conical temperature.

We  considered Reissner - Nordstrom black holes as well as some stringy
ones. In cases where the na\"{\i}vely defined temperature is finite, we found
the conical singularity argument to lead to the same value. On the other hand,
when the temperature defined from the surface gravity vanishes, as in the case
of the extremal Reissner - Nordstrom black hole, or blows up, as in the case
of the second extremal limit of the stringy
black hole considered by us, there is no genuine
conical singularity, and consequently no restriction on the temperature. In
the case of the extremal Reissner - Nordstrom black hole, we went on to
construct Kruskal-like coordinates corresponding to an arbitrary temperature,
but in the case of the stringy black hole considered above, the only coordinate
singularity is at $r=0$, and there is no question of defining new coordinates.

\acknowledgments
PM acknowledges the help provided by J. Maharana during the early stages of
this work at Bhubaneswar.

\end{document}